\documentclass[review]{elsarticle}

\usepackage{lineno,hyperref}
\modulolinenumbers[5]

\journal{Journal of \LaTeX\ Templates}

\begin{document}

\begin{frontmatter}

\title{Numerical study on a shelled gas bubble submerged in soft tissue}

\author[sahand]{F. Ghalichi}
\author[UUT]{S. Behnia\corref{mycorrespondingauthor1}}
\cortext[mycorrespondingauthor1]{Corresponding author}
\ead[UUT]{s.behnia@sci.uut.ac.ir}
\author[sahand]{F. Mottaghi}
\author[Bilkent]{M. Yahyavi\corref{mycorrespondingauthor2}}
 \cortext[mycorrespondingauthor2]{Corresponding author}
\ead[Bilkent]{m.yahyavi@bilkent.edu.tr}
\address[sahand]{Department of Mechanical Engineering, Division of Biomechanics,   Sahand University of Technology, Tabriz, Iran.}
\address[UUT]{Department of Physics, Urmia University of Technology, Urmia, Iran. }
\address[Bilkent]{Department of Physics, Bilkent University 06800, Ankara, Turkey.}

%

\begin{abstract}
Ultrasound contrast agents have been recently utilized in therapeutical implementations for targeted delivery of pharmaceutical substances. Radial pulsations of the encapsulated microbubbles under the action of an ultrasound field are complex and high nonlinear, particularly for drug and gene delivery applications with high acoustic pressure amplitudes. The dynamics of a polymer-shelled agent is inspected \textit{in vivo} through applying the method of chaos physics whereas the effects  of  the outer medium compressibility and the shell were considered. The stability of the ultrasound contrast agent is examined by plotting the bifurcation diagrams over a wide range of variations of influential parameters. The results imply that the composition of the surrounding medium alters the  microbubble dynamics, strongly. Furthermore, influences of various parameters which present a comprehensive view of the radial oscillations of the microbubble  are quantitatively discussed with clear descriptions of the stable and  unstable regions of the microbubble oscillations.\end{abstract}

\begin{keyword}
 Ultrasound contrast agent \sep Encapsulated microbubble \sep Targeted drug delivery\sep Nonlinear oscillation.\\
  \texttt{PACS: } 47.55.dd \sep 82.40.Bj \sep 95.10.Fh \sep 47.52.+j.
\end{keyword}

\end{frontmatter}

\section{Introduction}
\label{intro}
Ultrasound contrast agents (UCAs) are coated microbubbles by a stabilizing layer such as albumin, polymer or lipids whereas they are filled with a high-molecular-weight gas~\cite{1,2,3}. These agents are originally designed for diagnostic ultrasound imaging (for liver imaging, cardiac and other organs) since they are highly detectable with ultrasound imaging due to their great scattering properties, hence  this acoustic trait caused to the progression of more sensitive imaging methods~\cite{4,5}. Recently, in addition to diagnostic implementations, their employment in the biomedical field is promoting to the therapeutic applications~\cite{6} such as drug and gene delivery~\cite{7,8}, sonothrombolysis~\cite{9}, opening blood-brain barrier and delivery to the CNS~\cite{10,11,12}. Indeed, they are employed as transporters of pharmaceutical agents to carry them into the site of interest to spew their cargo just where it is needed by applying a focused ultrasound field~\cite{8,13,15,16}. This novel method has emerged immense clinical potentials such as minimizing drug-related toxicity to the healthy cells and tissues, drug dosage modifications, promoting transmembrane and extravascular drug transport, preventing drug-drug interactions, decreasing costs for the patient, additionally, transference and release can be visualized with real-time ultrasound and as a whole result treatment efficacy will be enhanced~\cite{17,18,19}.

UCAs undergo complex dynamic behaviors while they are exposed to an ultrasound field~\cite{20}. Depending on the applied acoustic amplitudes, the microbubble structure and the properties of the host media, they will respond   linear or nonlinear pulsations~\cite{21,22,23}. Fundamental perception of UCAs dynamics and precisely predicting their behavior will promote their diagnostic and therapeutic capabilities; indeed a quantitative understanding of UCAs dynamics is a necessary step to attain a better hardware design and successful clinical applications. Many sophisticated theoretical treatments for describing the coated microbubble response in an ultrasound field have been performed whereas most of the presented models are on the foundation of the Rayleigh-Plesset (RP) equation form. De Jong and co-workers~\cite{24,25} introduced the first a theoretical model that considers the encapsulation as a viscoelastic solid shell, as well as a damping coefficient term, is added to the RP equation. Church~\cite{26} presented a more accurate model by considering the shell thickness to describe the effects of the shell on UCA behavior. Morgan (see also Zheng)~\cite{27,28} and Allen~\cite{29} offered their models for thin and thick encapsulated microbubbles, respectively. Another rigorous model which treats the outer medium as a slightly compressible viscoelastic liquid is due to Khismatullin and Nadim~\cite{30}. Chatterjee and Sarkar~\cite{31} attempted to take account of the interfacial tension at the microbubble interface with infinite small shell thickness. Sarkar~\cite{32} improved this model to contain the surface elasticity by using a viscoelastic model. Stride and Saffari~\cite{33} demonstrated the presence of blood cells and the adhesion of them to the shell have a negligible effect. {Tamadapu and coworkers \cite{N1,N2} investigated an air-filled thick polymer encapsulated nonspherical microbubble  suspended in bulk volume of water.} Marmottant~\cite{34} exhibited a simple model for the dynamics of phospholipid-shelled microbubbles while taking account of a buckling surface radius, shell compressibility, and a break-up shell tension. Doinikov and Dayton~\cite{35} refined the church model and also considered the translational motion of the UCA. Shengping Qin and Katherine W. Ferrara~\cite{36} have presented a model to explain the radial oscillations of UCAs by considering the effects of liquid compressibility, the surrounding tissue, and the shell.
Although the aforementioned discussion expressed that considerable efforts have been performed, nonlinear dynamics of encapsulated microbubble by considering variations in different effective parameters is not fully realized by any means~\cite{37,38} and require supplemental developments. The nonlinear nature of the equation needs specialized tools for analyzing because linear and analytical solutions are inadequate. Based on the previous works, the chaotic behavior of free bubbles observed both theoretically and experimentally~\cite{39,40,411,412,413}, but this is not investigated for the case of UCAs, and it will be helpful to survey from this point of view because the method of chaos physics provides extensive knowledge about rich nonlinear dynamical systems. Moreover, neglecting liquid compressibility is not suitable for high-pressure amplitudes where the wall velocity of the agent is equal to the speed of sound in liquid~\cite{29}, so the effects of liquid compressibility on the microbubble dynamics should be considered~\cite{36,37}.

In this paper, the effects of substantial parameters that influence the UCA dynamics are studied in a large domain applying method of chaos physics and considering the compressibility of the outer medium and the shell. It will represent comprehensive information about extremely nonlinear pulsations of UCAs, particularly for drug and gene delivery applications where the applied acoustic pressure is considerably greater than the pressure employed in ultrasound imaging.

 \section{Mathematical model: Dynamics of a coated spherical microbubble}
The theoretical description of radial motion for a spherical encapsulated microbubble immersed in blood or tissue has been derived by Qin and Ferrara~\cite{36} which is utilized for numerical simulation. This justified equation also explains the effects of variations of the shell and the surrounding tissue on the UCA behavior and is given by:
 $$\left[\rho_L\left(1-\frac{\dot{R}_2}{c}\right)+\rho_S\left(1+\frac{\dot{R}_2}{c}\right) \left(\frac{R_2}{R_1}-1\right) \right] R_2\ddot{R}_2 +\left\{\frac{3}{2}\rho_L \left(1-\frac{1}{3}\frac{\dot{R}_2}{c}\right)+\right.$$
 $$
\left. \rho_S \left(1+\frac{\dot{R}_2}{c}\right)\right. \left.  \left[-\frac{3}{2}+2 \left(\frac{R_2}{R_1}\right)-\frac{1}{2}\left(\frac{R_2}{R_1}\right)^4\right] \right\}\dot{R}_2^2=\left(1+\frac{\dot{R}_2}{c}\right)\left \{ p_g(t)-\frac{2\sigma_1}{R_1}- \right.$$
$$\frac{2\sigma_2}{R_2}-\frac{4}{3}G_S\left[1-\left(\frac{R_{20}}{R_2}\right)^3\right]\frac{R_S}{R_2^3-R_S}-
4\mu_S\frac{R_S}{R_2^3-R_S}\frac{\dot{R}_2}{R_2}-\frac{4}{3}G_L \left[1-\left(\frac{R_{20}}{R_2}\right)^3\right]-$$
\begin{equation}\label{Microbubble}{}
  \hspace{-4cm} \left.4\mu_L\frac{\dot{R}_2}{R_2}-p_0-
p_i(t)\right\}-3\gamma\frac{\dot{R}_2}{c}
\left(\frac{R_2}{R_{10}}\right)^3\frac{p_g(t)}{\left(\frac{R_1}{R_{10}}\right)^3-\frac{b}{V_m}}.
\end{equation}
\textbf{where $p_{i}(t)=P_a \sin(2\pi f t)$ is the ultrasound pressure at infinity. Also, the pressure-volume
relation, $p_g$,  is defined as follows}
\begin{equation}
p_g(t)=\left(p_0+\frac{2\sigma_1}{R_{10}}+\frac{2\sigma_2}{R_{20}}\right)\left[\frac{1}{\left(\frac{R_1}{R_{10}}\right)^3-\frac{b}{V_m}}\right]^\gamma
\end{equation}
 {The equation (\ref{Microbubble}) is applied to describe nonlinear oscillations of a polymer-shelled agent versus variations of several important parameters. This model was developed to describe the dynamics of UCAs in \textit{vivo} while taking account of the effects of the surrounding tissue, the shell tissue, and liquid compressibility. In the literature, the correction term for compressibility has different forms for different considerations. In this work,  we choose the form $(R/c) (dp_g(t) /dt)$ as in Ref. \cite{NN1}. Since the time derivative of the driving pressure $(dp_g(t) /dt)$ is small and not dominant for violent oscillation. The assumption used in this paper is that the shell thickness is finite and the shell material behaves as a Voigt viscoelastic solid. The Qin-Ferrara is similar to the Church model \cite{26}, but, the Church shell elastic term is valid only for small deformation since in the Qin-Ferrara's model, the shell elastic term is stated to be valid for finite deformation of the shell.
}
\subsection{Variables and its domain}
 {The evolution of microbubbles dynamics corresponds to the different parameter, which should be explained separately. As the inner and outer radius of the agent is described with $R_{1}$ and $R_{2}$, then  $\dot{R}_{1}$ and $\dot{R}_{2}$ are the inner and outer wall velocity of the agent, respectively. $\ddot{R}_{2}$ is the outer wall acceleration of the agent. Naturally, $R_{10}$ and $R_{20}$ used as initial outer and inner radius of bubbles. Also, shell thickness is chosen as $R_S=R_{20}^3-R_{10}^3$. $\rho_{L}$ is the density of the liquid and $\rho_{S}$ is the Shell density. $c$ is the speed of sound in the liquid.
$\sigma_1$ is the inner surface tension, $\sigma_2$ is the outer surface tension, $p_{g}$ is the gas pressure within the agent. $G_{S}$ is the shear modulus, $G_{L}$ is the shear modulus of the surrounding medium which represent the stiffness of the surrounding tissue.
$\mu_{L}$  is the viscosity of the liquid, $\mu_{s}$ is shell viscosity.
 Finally, $f$ is the ultrasound center frequency, $b$ is the van der Waals constant, $V_{m}$ is the universal molar volume, $p_{0}$ is the hydrostatic pressure.
The introduced constants and varied parameter values for polymer-shelled agent are summarized in Tables \ref{table:1} and \ref{table:22}~\cite{23,33}.}

\begin{table}[h!]
\centering
\begin{tabular}{cccc}
\hline
\ Symbol&Parameter&Value&Unit\\
\hline
$\rho_{S}$ &Shell density &1150 &$\frac{kg}{m^{3}}$\\
$\rho_{L}$ &Liquid density &1060 &$\frac{kg}{m^{3}}$\\
$\mu_{L} $&Liquid viscosity&0.015&Pa s  \\
$\sigma_{1}$ &Surface tension at inner radius &0.04&N $m^{-1}$\\
$\sigma_{2}$  &Surface tension at outer radius&0.056&N $m^{-1}$\\
$R_{10}$&Equilibrium inner radius of agent&2.3750& $\mu$m \\
$R_{20}$&Equilibrium outer radius of agent&2.5&$\mu$m  \\
$p_{0}$&Hydrostatic pressure&1.01&$\times10^{5}$ Pa\\
$c$ &Sound speed in liquid &1540 &$\frac{m}{s}$\\
$b$ &Van der Waals constant&0.1727& $\frac{l}{mol}$\\
$V_{m}$&Universal molar volume&22.4& $\frac{l}{mol}$\\
$\gamma$ & Polytropic gas exponent &1.4&\\
\hline
\end{tabular}
\caption{Physical constants parameters for polymer-shelled agent~\cite{23,33}. }
\label{table:1}
\end{table}

\begin{table}[h!]
\centering
\begin{tabular}{cccc}
\hline
\ Parameter& Range of value&Unit\\
\hline
Driving pressure& $0<P_a<2$&Pa \\
Driving frequency& $0.5<f<5$&Hz \\
Shear modulus of surrounding medium&$0<G_{L}<1.5$&MPa\\
Shell viscosity&$0<\mu_{S}<5$&Pa s  \\
Shear modulus of shell&$0<G_{S}<200$ &MPa\\
Shell thickness& $0<R_S$ $<0.15$&$\mu$m\\
\hline
\end{tabular}
\caption{Physical  varied  parameters for polymer-shelled agent~\cite{23,33}. }
\label{table:22}
\end{table}

\section{Results}
\label{Results}
\begin{figure}
\begin{center}
\resizebox*{\linewidth}{!}{\includegraphics{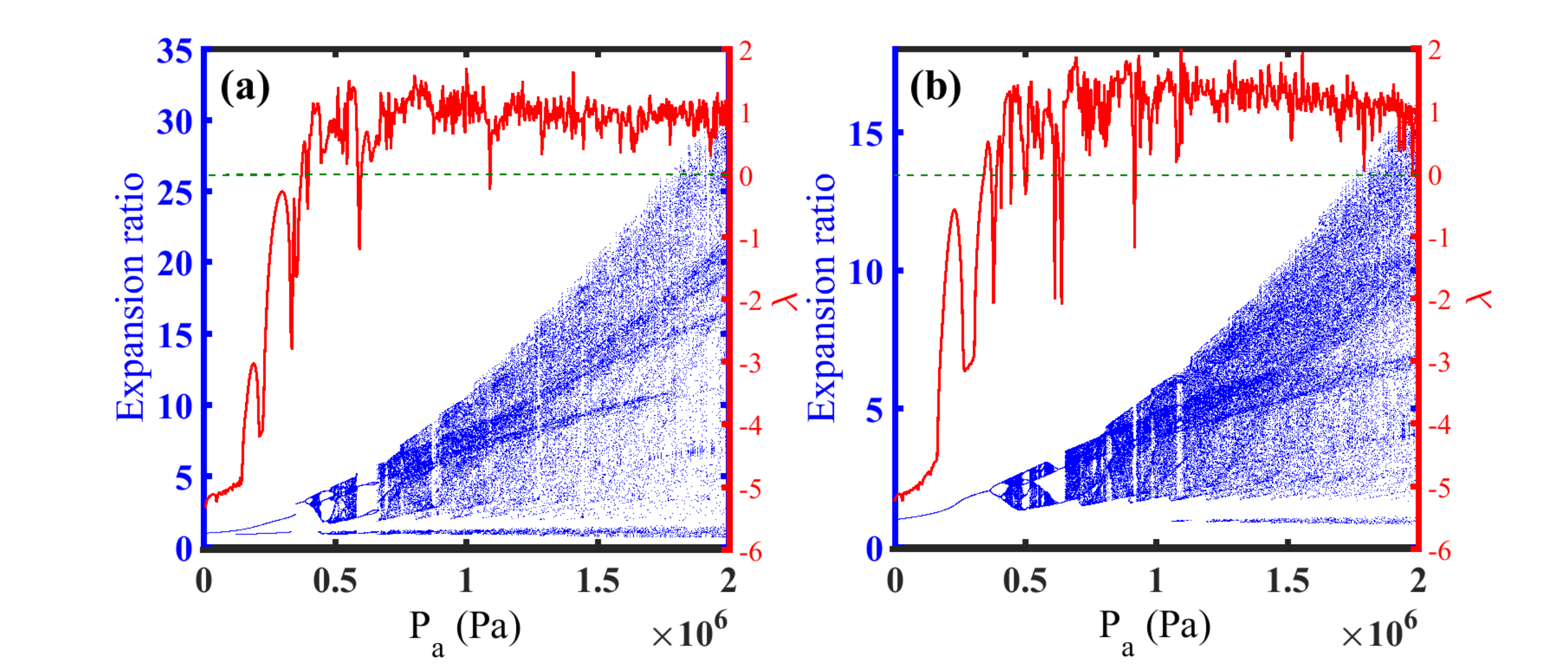}}\\%
\resizebox*{\linewidth}{!}{\includegraphics{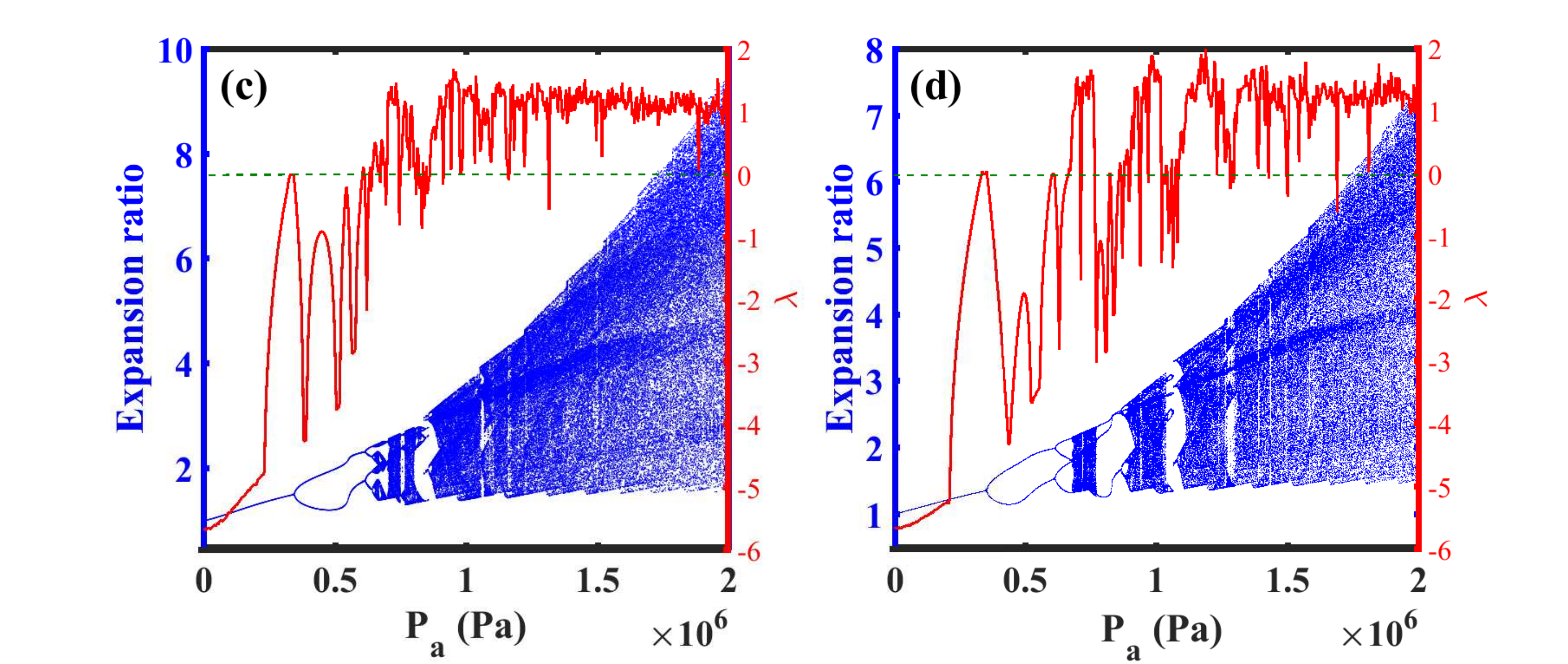}}\\
\resizebox*{\linewidth}{!}{\includegraphics{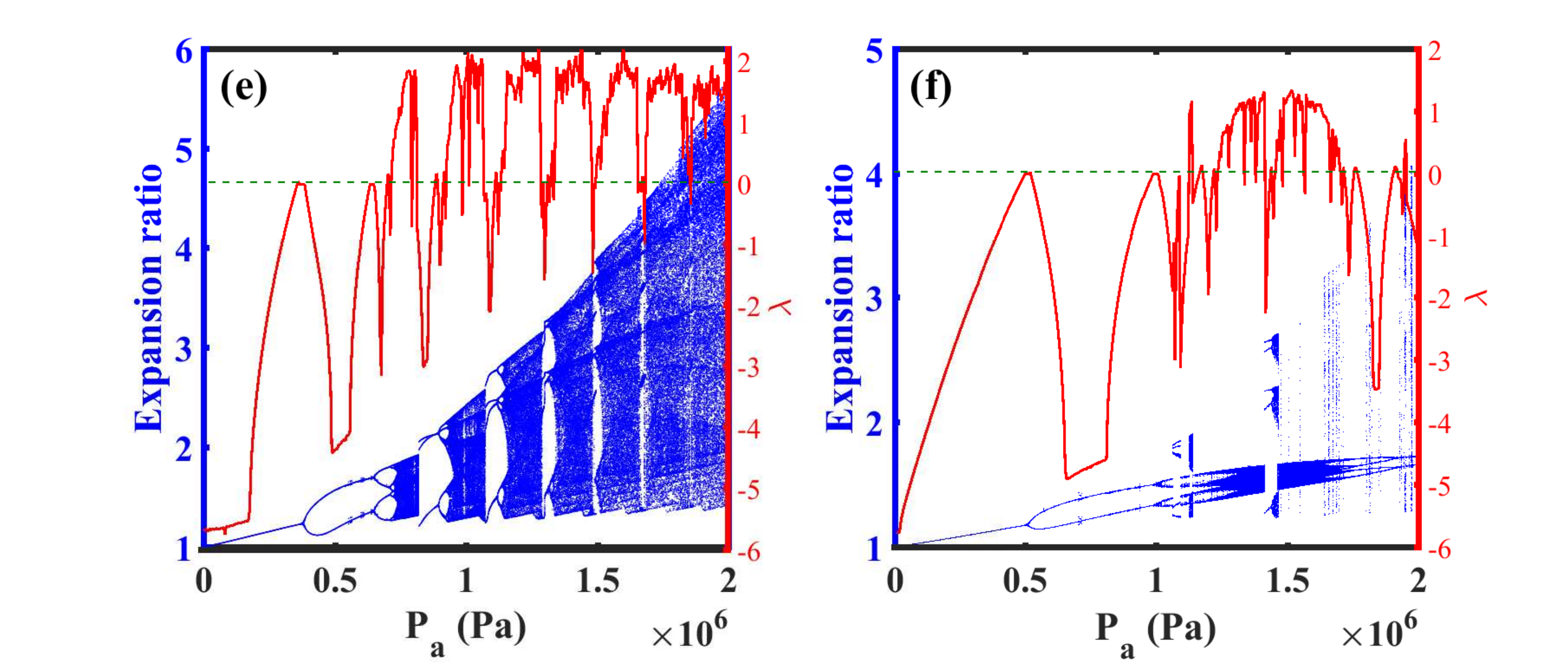}}
\caption{ Bifurcation diagrams (Expansion ratio-blue dot points) and Lyapunov exponent ($\lambda$-red solid line) of normalized microbubble radius versus driving pressure whit $G_L=0$ MPa, $\mu_s=0.45$ Pa s, and $G_S=11.7$ MPa while the frequency is (a) 0.6 MHz, (b) 1 MHz, (c) 1.5 MHz, (d) 1.8 MHz, (e) 2.2 MHz and (f) 2.8 MHz. All other physical parameters were kept constant at values given in Table \ref {table:1}.}%
\label{fig1}
\end{center}
\end{figure}

For a better visualization of the evolution of the effect of acoustic pressure alterations on microbubble dynamics, radial motion of a UCA is investigated versus a prominent domain of acoustic pressure from 10 kPa to 2 MPa. Fig. \ref{fig1}a-f show the bifurcation diagrams and Lyapunov exponent ($\lambda$) of the normalized microbubble radius against acoustic pressure as the control parameter for several values of applied frequency of the ultrasound field which they are 0.6, 1, 1.5, 1.8, 2.2 and 2.8 MHz, respectively. In each one stable and chaotic pulsations can be observed,  regarding the sign of the corresponding Lyapunov exponent. The existence of the negative (positive) Lyapunov exponent indicates the stable (chaotic) behavior. It is perceived that by raising pressure amplitudes the microbubble stability is reduced and  chaotic oscillations will be evident which this trend can be confirmed by other works~\cite{34,42,43}. Regarding Fig. \ref{fig1}a-f, the microbubble experiences distinctive behaviors in different amplitudes of frequency while the control parameter (acoustic pressure) is increasing. As it is followed in Fig. \ref{fig1}a-f, by amplifying the magnitude of frequency the microbubble stability is enlarged in superior driving pressures which is in a good agreement with the work of~\cite{44}. This is obvious in Fig. \ref{fig1}f, where the accessibility to the stable range concerning variations of pressure has the maximum extent.

According to Fig. \ref{fig1}d, the radial motion of the microbubble in frequency 1.8 MHz manifests stable behavior of period one till 352 kPa which is followed by a period doubling up to 615 kPa, after that the system demonstrates a period four for a small interval and it is pursued by the first chaotic window in 680 kPa ($\lambda>0$). Then the microbubble exhibits its periodic behavior again before the next jump to chaos in 867 kPa. These intermittent transitions between chaotic oscillations and stable behavior persist until 1.3 MPa, afterward, the system turns into severely chaotic oscillations which continues to the termination of pressure interval, i.e., 2 MPa. This behavior is also observed experimentally in~\cite{45}. The microbubble experience more stability in a broad domain of driving pressure and chaotic pulsations and the expansion ratio of the UCA is reduced while the applied frequency is higher (see Fig. \ref{fig1}f).

\begin{figure}
\begin{center}
\resizebox*{\linewidth}{!}{\includegraphics{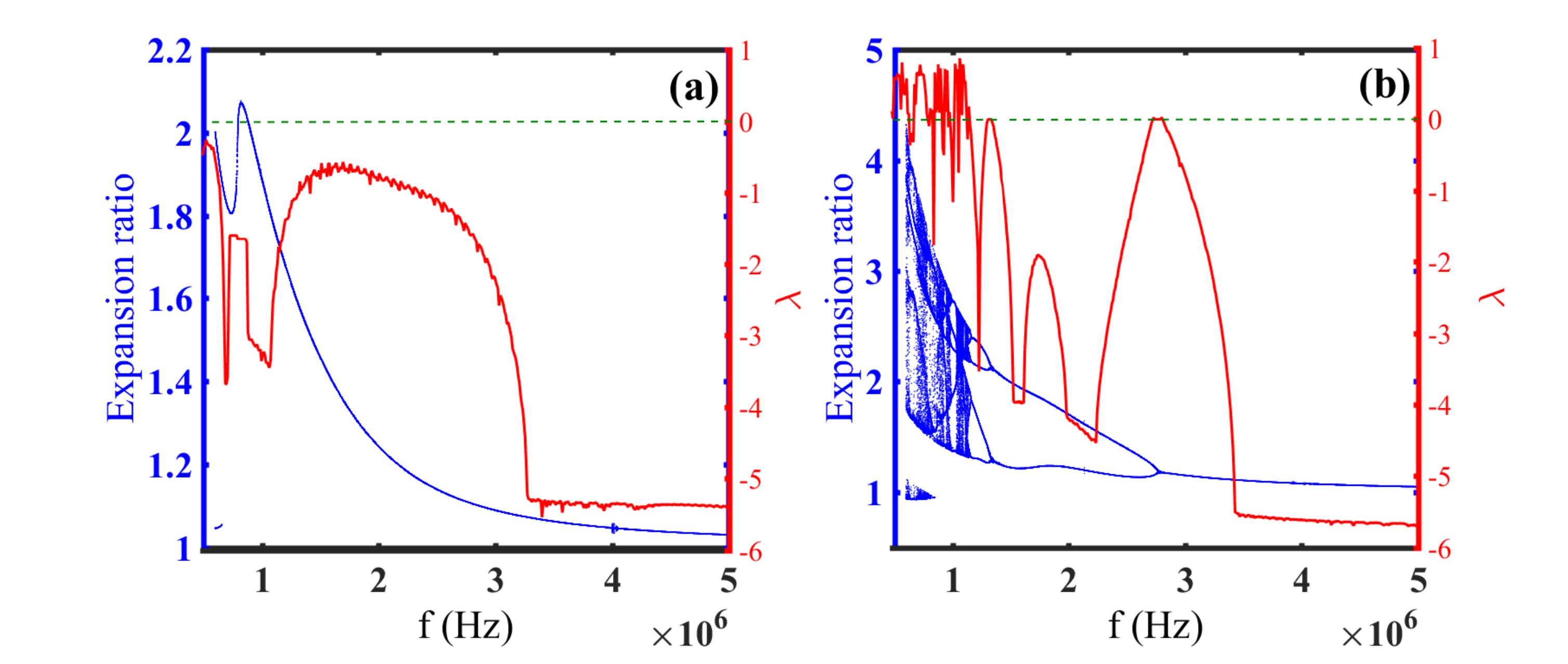}}\\%
\resizebox*{\linewidth}{!}{\includegraphics{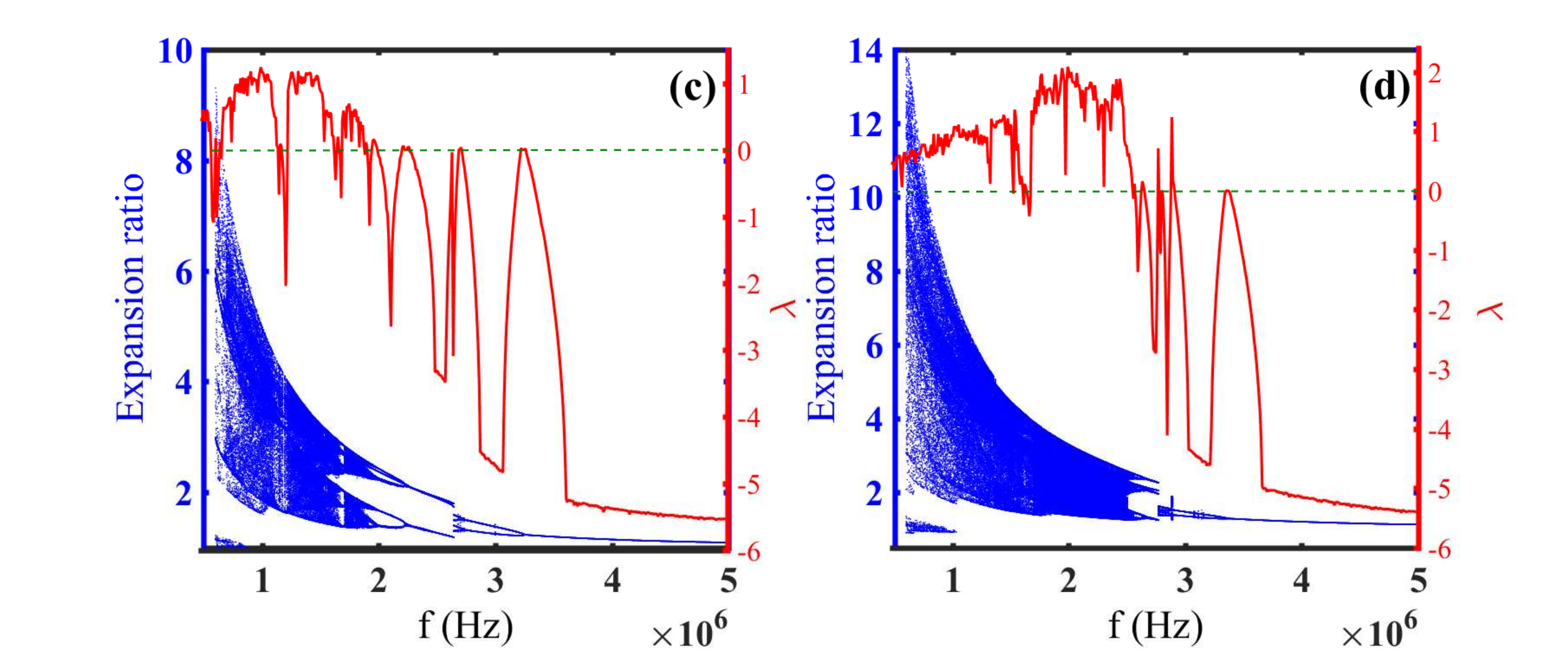}}\\
\resizebox*{\linewidth}{!}{\includegraphics{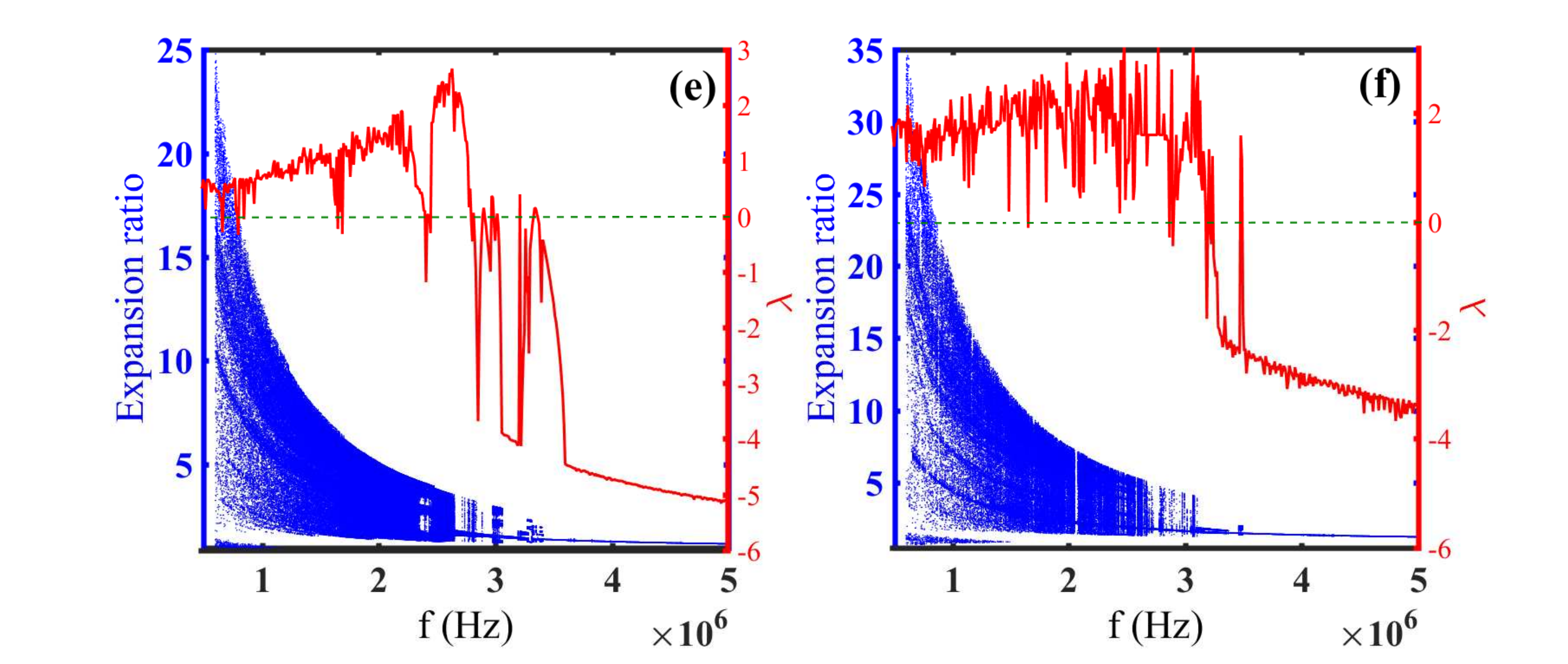}}
\caption{Bifurcation diagrams (Expansion ratio-blue dot points) and Lyapunov exponent ($\lambda$-red solid line) of normalized microbubble radius versus driving frequency  whit $G_L=0$ MPa, $\mu_s=0.45$ Pa s, and $G_S=11.7$ MPa while the acoustic pressure is (a) 0.3 MPa, (b) 0.5 MPa, (c) 0.9 MPa, (d) 1.2 kPa, (e) 1.7 MPa and (f) 2.2 MPa.}%
\label{fig2}
\end{center}
\end{figure}

Also for studying the effect of frequency alterations on microbubble dynamics, the dynamical behavior of UCA is inspected by considering the ultrasound frequency as the control parameter which varies from 600 kHz to 5 MHz, the corresponding bifurcation diagrams and Lyapunov exponent  of the normalized microbubble radius is shown in Fig. \ref{fig2}a-f for the applied pressure values of 0.3, 0.5, 0.9, 1.2, 1.7, 2.2 MPa, respectively.

The stable behavior of microbubble is presented for the low amplitude of pressure ($\lambda<0$), i.e., 0.3 MPa (Fig. \ref{fig2}a). The chaotic behavior ($\lambda>0$) of UCA appears by increasing the values of applied pressure (Fig. \ref{fig2}b), and the microbubble shows more chaotic oscillations as the pressure is intensifying (Fig. \ref{fig2}b-f) which this phenomenon is seen in~\cite{22}. It is seen in all figure \ref{fig2}a-f that, the magnitude of pulsations reduces significantly and the chaotic region becomes smaller when the control parameter (frequency) is increasing, and the UCA shows the stable behavior of period one which reveals the stabilizing property of superior frequencies which is confirmed in~\cite{29}. It is seen in all of them (Fig. \ref{fig2}a-f) that UCA goes to stable manner at high values of frequency and microbubbles stimulating with superior pressures become stable at superior frequencies.

\begin{figure}
\begin{center}
\resizebox*{\linewidth}{!}{\includegraphics{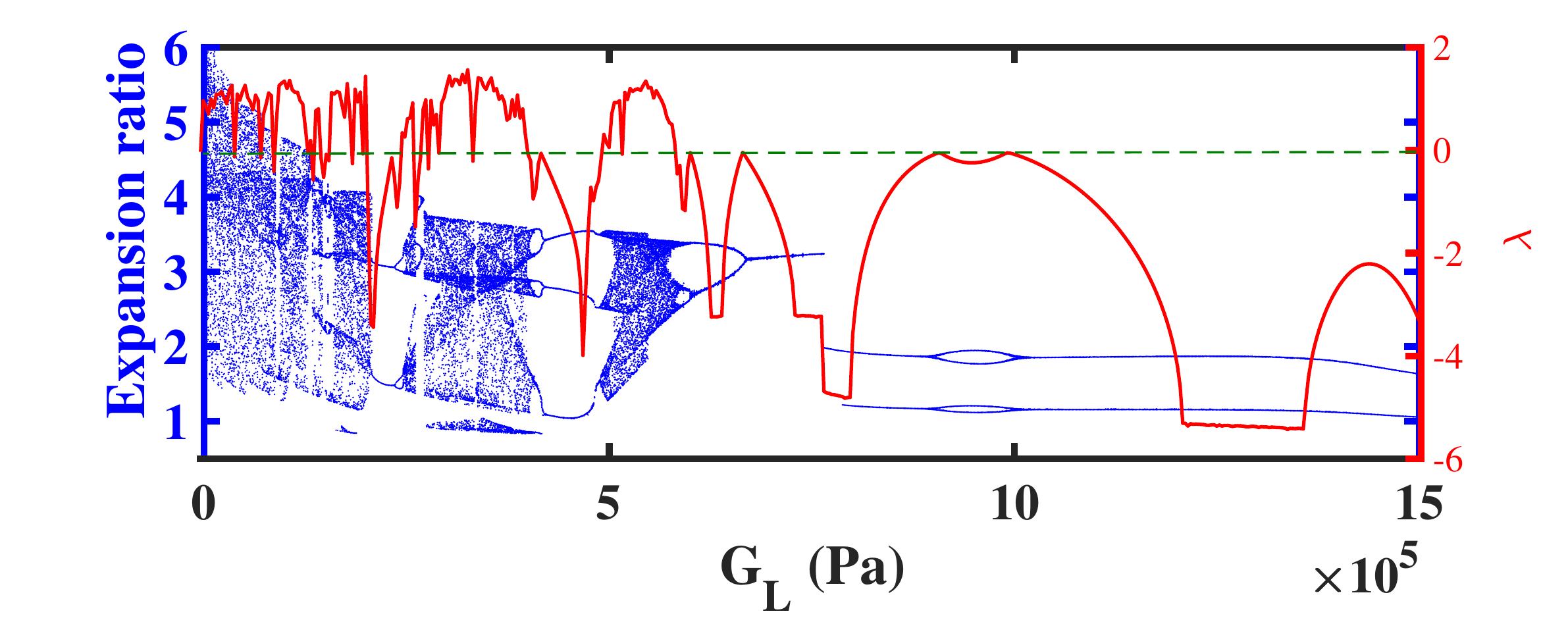}}
\caption{Bifurcation diagrams (Expansion ratio-blue dot points) and Lyapunov exponent ($\lambda$-red solid line) of normalized microbubble radius versus shear modulus of surrounding medium whit $\mu_s=0.45$ Pa s, and $G_S=11.7$ MPa when the driving frequency and pressure are, respectively, 1.5 MHz and 1.5 MPa. }%
\label{fig3}
\end{center}
\end{figure}
\begin{figure}
\begin{center}
\resizebox*{\linewidth}{!}{\includegraphics{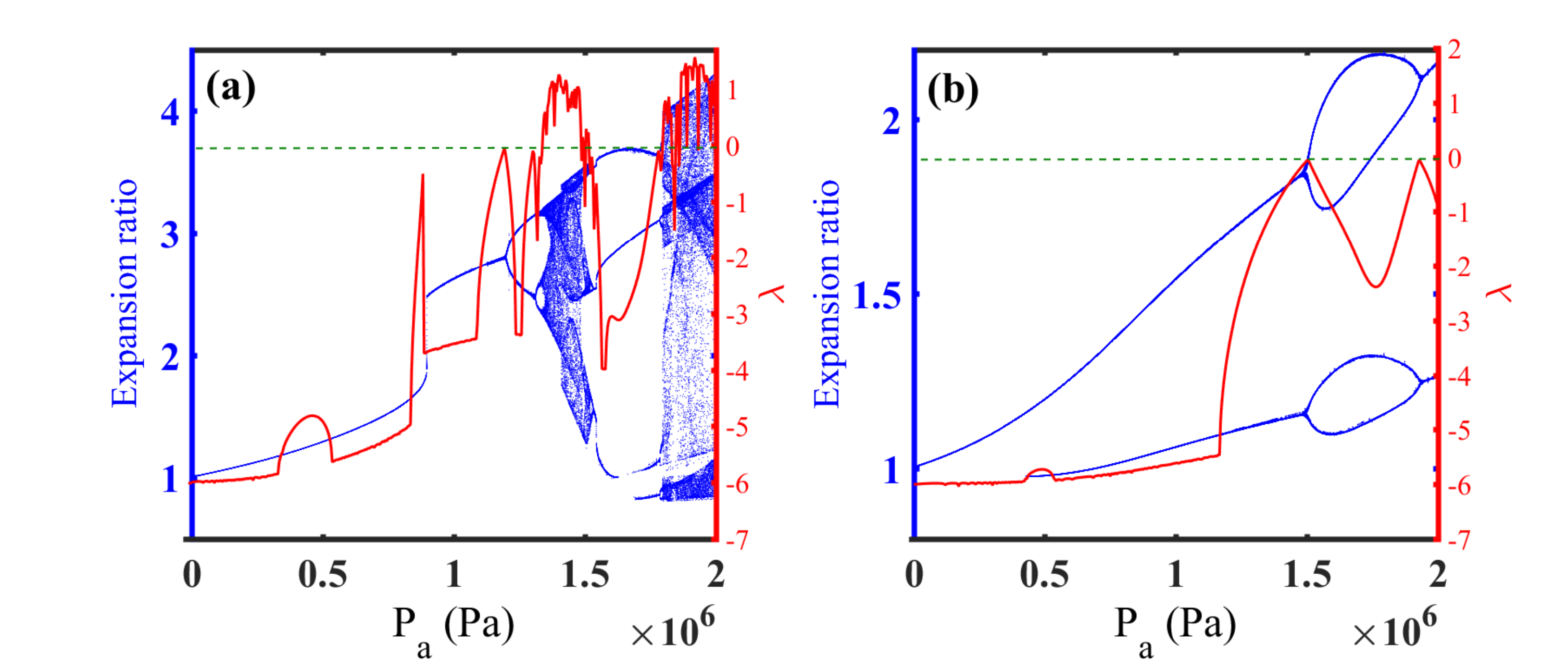}}
\caption{Bifurcation diagrams (Expansion ratio-blue dot points) and Lyapunov exponent ($\lambda$-red solid line) of normalized microbubble radius versus driving pressure whit $\mu_s=0.45$ Pa s, and $G_S=11.7$ MPa  when the applied frequency is 1.5 MHz for the surrounding medium with $G_L$ (a) 0.5 MPa, (b) 1 MPa.}%
\label{fig4}
\end{center}
\end{figure}
\begin{figure}
\begin{center}
\resizebox*{\linewidth}{!}{\includegraphics{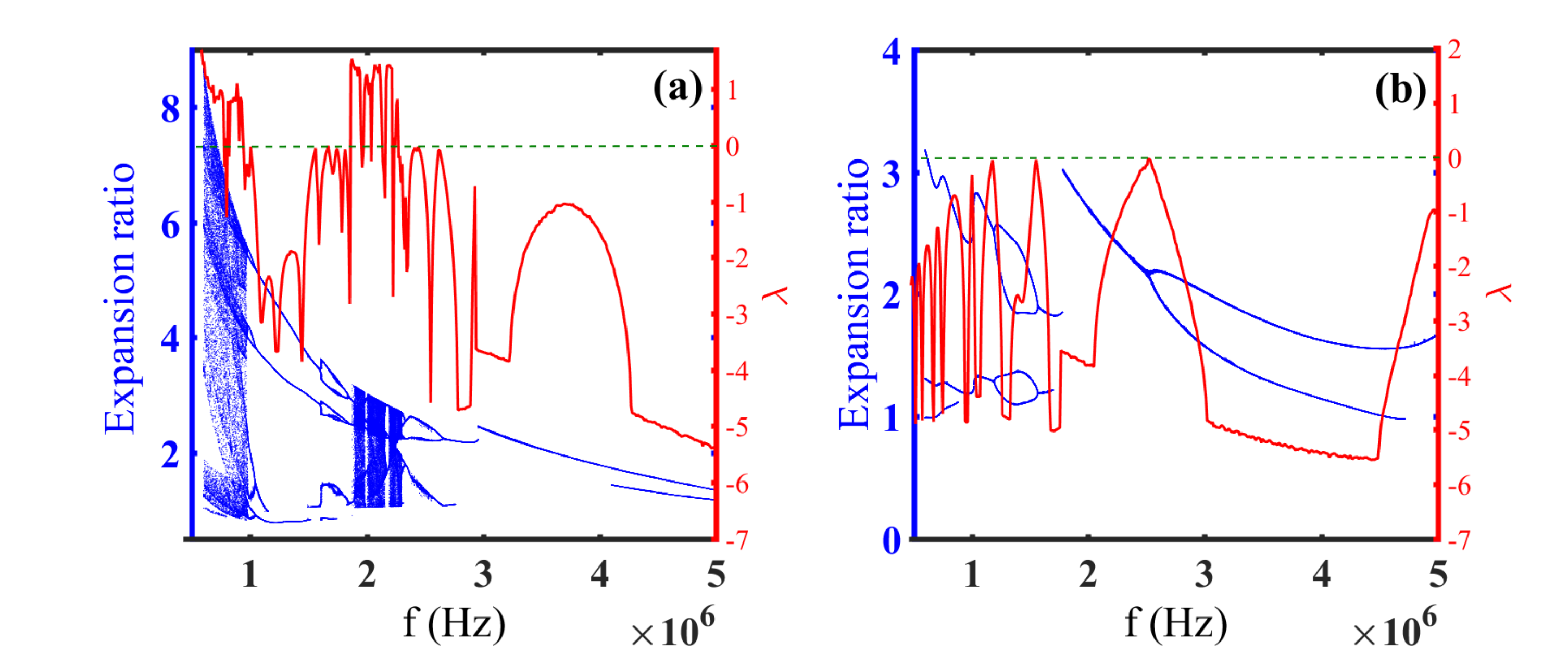}}
\caption{Bifurcation diagrams (Expansion ratio-blue dot points) and Lyapunov exponent ($\lambda$-red solid line) of normalized microbubble radius versus driving frequency whit $\mu_s=0.45$ Pa s, and $G_S=11.7$ MPa when the acoustic pressure is 1.7 MPa for the surrounding medium with $G_L$ (a) 0.5 MPa, (b) 1 MPa.}%
\label{fig5}
\end{center}
\end{figure}

The mechanical characteristics of the medium that surrounds the UCA are  varied with kind of tissue and its composition. Therefore, the effects of shear modulus of tissue  on microbubble behavior are studied by considering shear modulus variations from 0 to 1.5 MPa. Bifurcation diagram  and Lyapunov exponent  of normalized microbubble radius are demonstrated by taking the shear modulus of the surrounding medium as the control parameter (Fig. \ref{fig3}) whereas the acoustic pressure amplitude is 1.5 MPa and the ultrasound frequency is 1.5 MHz. Moreover, bifurcation diagrams  and Lyapunov exponent  of normalized microbubble radius versus acoustic pressure (Fig. \ref{fig4}a-b) and applied frequency (Fig. \ref{fig5}a-b) are plotted within the soft tissue and proportionately stiff tissue with $G_L$=0.5 and 1 MPa, respectively.

It is observed that the microbubble behavior is chaotic for low values of shear modulus (Fig. \ref{fig3}) and as the shear modulus of the outer medium is increasing the expansion ratio of the microbubble and chaotic pulsations are reducing. The microbubble finally goes to the stability by increasing the magnitude of shear modulus of the surrounding medium up to 765 kPa which can be confirmed in~\cite{36}.

In the same conditions, the microbubble behavior is probed versus acoustic pressure for two different values of the shear modulus of the surrounding medium, i.e., 0.5 and 1 MPa (see Fig. \ref{fig4}a-b). Comparing these figures in Fig. \ref{fig1}c reveals that the oscillations abate by increasing the magnitude of the shear modulus of the medium; indeed the system has more stability when the external medium is more rigid. The microbubble demonstrates various dynamical behaviors in 3 values of $G_L$. When the microbubble is surrounded by blood with $G_L$=0 (see Fig. \ref{fig1}c), it undergoes more chaotic oscillations in lower pressure amplitudes, e.g., the first chaotic window is indicated in 677 kPa, this incident takes place in 1.27 MPa beside soft tissue (Fig. \ref{fig4}a) and the system is completely stable for the case of comparatively hard tissue with $G_L$=1 MPa (Fig. \ref{fig4}b).

The effects of frequency variations on microbubble behavior for the values of $G_L$=0.5, 1 MPa are plotted when the acoustic pressure is 1.7 MPa (Fig. \ref{fig5}a-b). Comparing these results with Fig. \ref{fig2}e which shows the effect of frequency variations in $G_L$=0 manifests this fact that by increasing the magnitude of shear modulus of the surrounding medium, the chaotic oscillations decrease significantly and as it is seen in (Fig. \ref{fig5}b)  the chaotic behavior disappears for $G_L$=1 MPa. It is also evident that the expansion ratio of the microbubble is smaller for higher magnitudes of $G_L$, in fact, the nonlinearity intensifies for smaller values of $G_L$.
\begin{figure}
\begin{center}
\resizebox*{\linewidth}{!}{\includegraphics{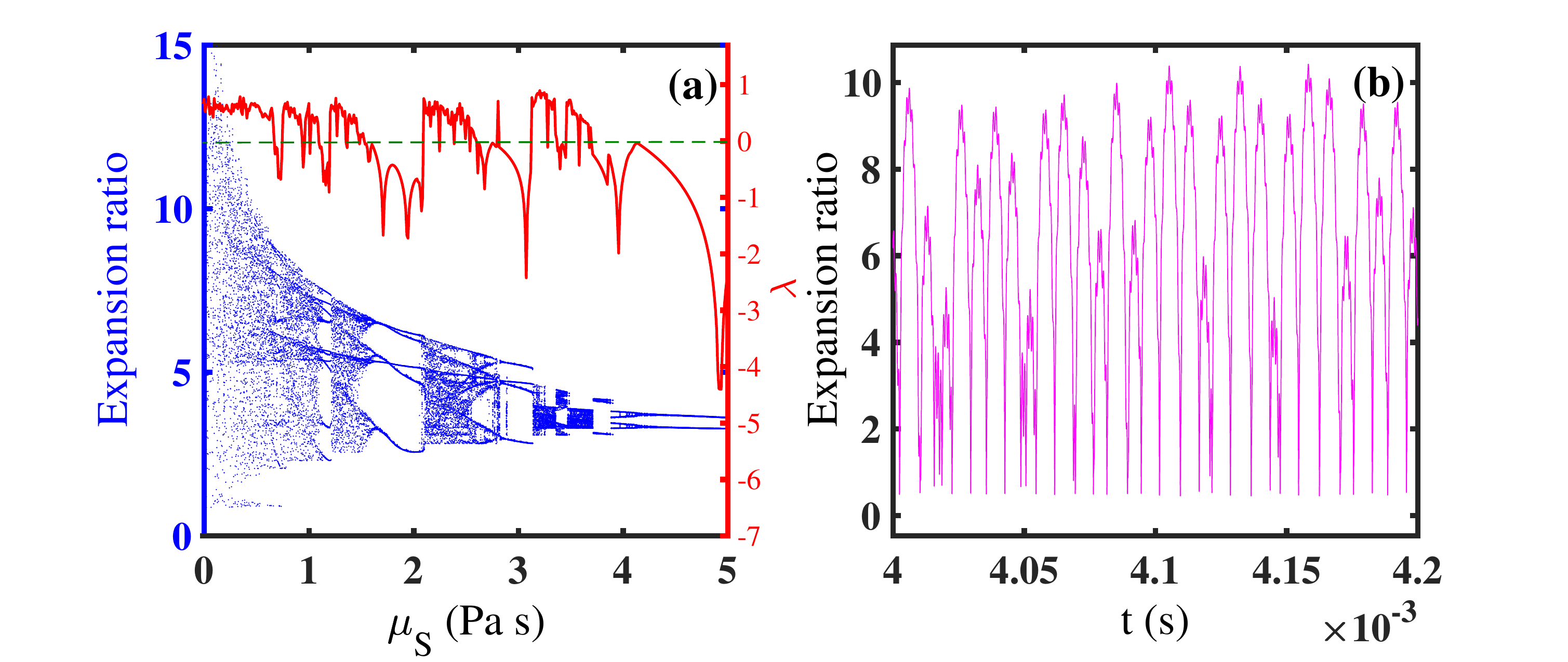}}
\caption{ (a) Bifurcation diagrams (Expansion ratio-blue dot points) and Lyapunov exponent ($\lambda$-red solid line) of normalized microbubble radius versus shell viscosity when the driving frequency and pressure are, respectively, 1 MHz and 1.5 MPa for $G_L$=0, and $G_S=11.7$ MPa, (b) The corresponding time series of normalized oscillations with the shell viscosity $\mu_s=0.45$ Pa s.}
\label{fig6}
\end{center}
\end{figure}

During our investigation, to explore the effect of shell viscosity alterations on microbubble dynamics,
bifurcation diagrams and Lyapunov exponent of normalized microbubble radius are plotted versus shell viscosity as the control parameter. Its value varies in the range 0.01 to 5 Pa s for 3 values of $G_L$ (0, 0.5, 1 MPa) in frequency=1 MHz and acoustic pressure=1.5 MPa.\\
 Results represent that the expansion ratio of the microbubble is much higher for low values of shell viscosity in $G_L$=0 (Fig. \ref{fig6}a) and also it is evident that by increasing the value of shell viscosity the nonlinearity and the maximum microbubble expansion decrease which is seen in~\cite{26,30,47}. Fig. \ref{fig6}b demonstrates the normalized oscillations of the microbubble versus time in frequency=1 MHz and acoustic pressure=1.5 MPa when the microbubble is surrounded by blood with $G_L$=0. This figure represents the chaotic oscillations of the microbubble for a definite value of the shell viscosity, i.e., 0.45 Pa s and as it is seen the maximum expansion ratio in this value is the same as Fig. \ref{fig6}a.

The microbubble exhibits fully chaotic behavior for small values of shell viscosity which is pursued by period doubling and the system reaches to period one stability in 0.66 Pa.s for $G_L$=0.5 MPa. The UCA dynamics is completely stable in $G_L$=1 MPa when the UCA is surrounded by relatively stiff tissue (results was not shown here).

\begin{figure}
\begin{center}
\resizebox*{\linewidth}{!}{\includegraphics{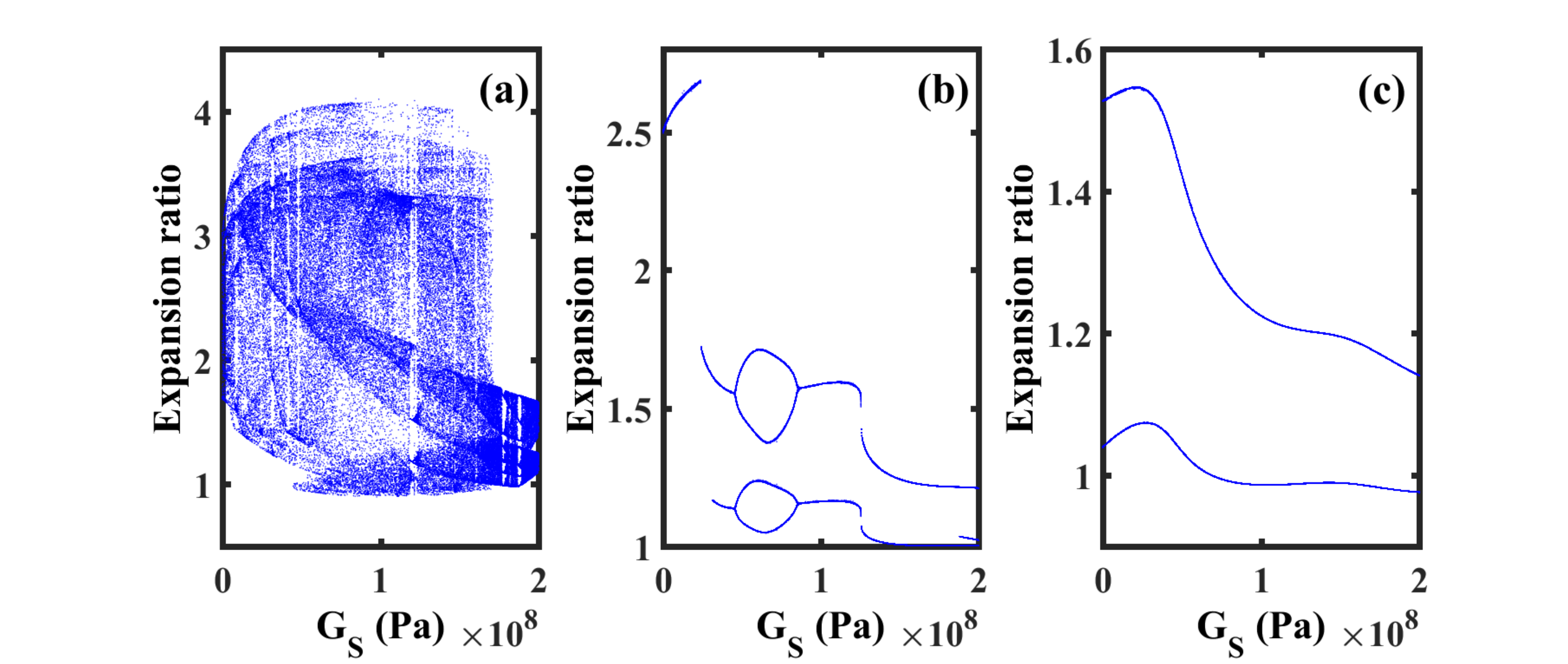}}
\caption{Bifurcation diagrams of normalized microbubble radius versus shear modulus of shell when the driving frequency and pressure are, respectively, 1.5 MHz and 1 MPa for $\mu_s=0.45$ Pa s,  $G_S=11.7$ MPa, and $G_L$ is (a) 0 MPa, (b) 0.5 MPa, (c) 1 MPa.} \label{fig7}
\end{center}
\end{figure}
\begin{figure}
\begin{center}
\resizebox*{\linewidth}{!}{\includegraphics{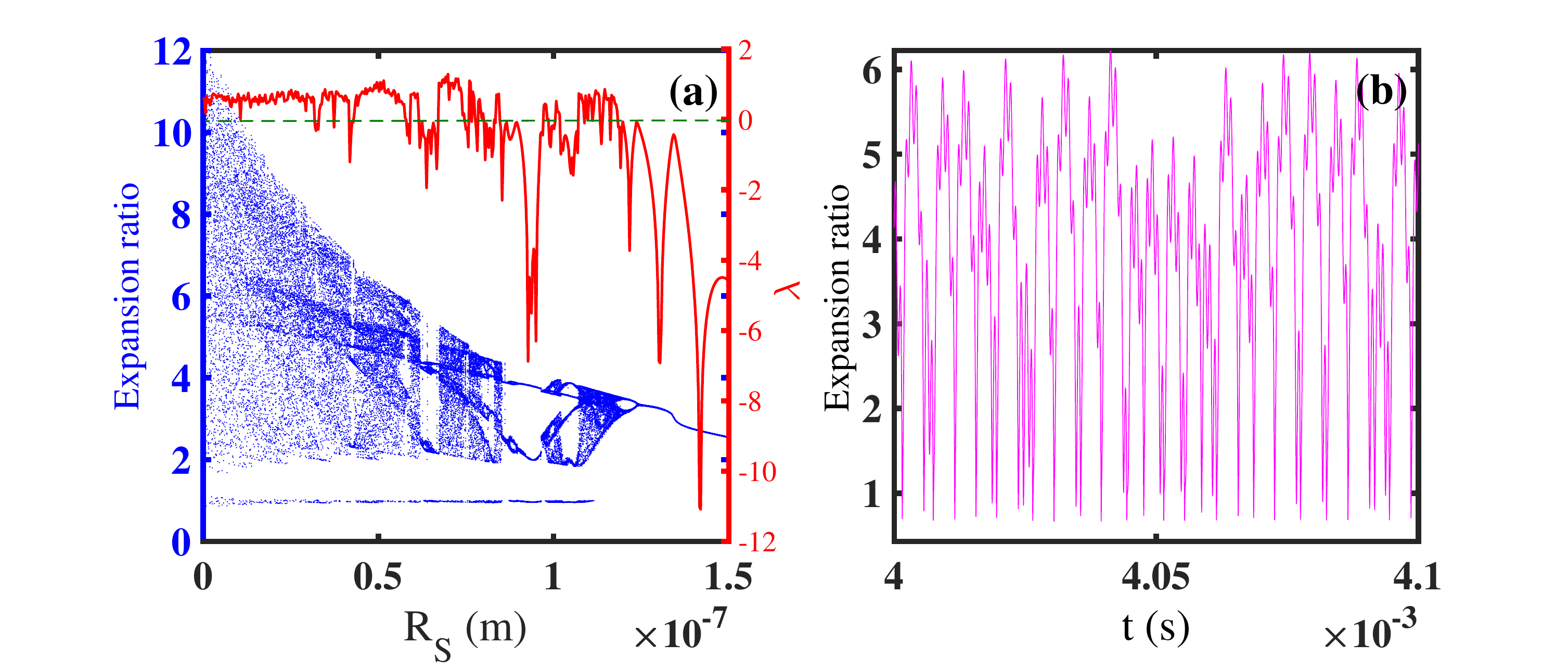}}
\caption{(a) Bifurcation diagrams (Expansion ratio-blue dot points) and Lyapunov exponent ($\lambda$-red solid line) of normalized microbubble radius versus shell thickness when the driving frequency and pressure are, respectively, 1 MHz and 1.5 MPa for $\mu_s=0.45$ Pa s,  $G_S=11.7$ MPa, and $G_L$=0, (b) The corresponding time series of normalized oscillations with the shell thickness. R$_S=50$ nm.}
\label{fig8}
\end{center}
\end{figure}

Next, by employing the values of 1.5 MHz and 1 MPa for driving frequency and pressure, respectively, but this time  considering the shear modulus of the shell as the control parameter while varying between 0 to 200 MPa, the bifurcation diagrams are presented for 3 values of $G_L$ (0, 0.5, 1 MPa).\\
By Fig. \ref{fig7}a-c, the microbubble response is entirely disparate in $G_L$=0 (Fig. \ref{fig7}a) with regards to $G_L$=0.5 MPa (Fig. \ref{fig7}b) and $G_L$=1 MPa (Fig. \ref{fig7}c). Its dynamics is stable for the values of $G_L$=0.5 and 1 MPa (Fig. \ref{fig7}b-c) while it exhibits chaotic oscillations and high expansion ratio in $G_L$=0 and the chaotic region becomes narrower in $G_S$=171 MPa (Fig. \ref{fig7}a).

One of the most important parameters that influence the microbubble behavior is the shell thickness of the microbubble which is utilized as the control parameter and varying in the range of 0 to 150 nm with the values of frequency and pressure of 1 MHz and 1.5 MPa, respectively. The bifurcation diagrams are sketched for three  values of $G_L$ (0, 0.5, 1 MPa).

Fig. \ref{fig8}a exposes that the UCA endures chaotic pulsations in a small magnitude of the shell thickness and increasing the shell thickness decreases the expansion ratio of the microbubble diameter and the system becomes stable when the shell thickness of the agent is 124 $nm$. These results approve the previous works in a very wide range of shell thickness variations~\cite{26,30,29}. Fig. \ref{fig8}b presents the corresponding time series of the normalized oscillations of the microbubble in frequency=1  MHz  and acoustic pressure=1.5  MPa  while the shell thickness of the agent is 50 nm and the microbubble is surrounded by blood with $G_L$=0. It is evident that the amplitude of pulsations is the same value as Fig. \ref{fig8}a.

For 0.5 MPa of $G_L$ the microbubble dynamics is chaotic for a small interval of shell thickness up to 18 nm and goes to the stable manner which lasts as a predominant situation to the end of the interval. For 1 MPa of $G_L$ the microbubble remains stable in the whole range of shell thickness and any chaotic behaviors is not viewed (results was not shown here).

The stable domains of the polymer-shelled agent are summarized in Table \ref{table:2}  for some consequential parameters. These results reveal that the stiffness, of the surrounding medium influences the UCA behavior impressively and also demonstrates the chaotic oscillations of UCA under the action of an ultrasound field which can be used to distinguish stable and unstable regions of microbubble pulsations and the expansion ratio of the UCA.

\begin{table}[h!]
\centering
\begin{tabular}{|c||ccc||r|}
	\hline
Parameter  &  $G_L=0$&   $G_L=0.5$& $G_L=1$& Unit    \\
	\hline \hline
Pressure   &$<0.65$  &$<1.3$  &entirely stable  &MPa \\
Frequency   &$>2.65$  &$>2.45$  &entirely stable  &MHz\\
Shell viscosity   &$>2.82$  &$>0.53$  &entirely stable  &Pa s   \\
Shell thickness   &$>124$  &$>18$  &entirely stable  &nm \\
Shear modulus of shell   &$>171$  &entirely stable  &entirely stable  &MPa \\
	\hline
\end{tabular}
\caption{Stable regions of polymer-shelled agent versus variations of various parameters in $G_L$=0 (blood), 0.5 (soft tissue) and 1 (stiff tissue) MPa. }
\label{table:2}
\end{table}

\section*{Conclusions}
\label{Conclusions}
{This article explained the dynamics driven a shelled gas bubble submerged in soft tissue by using the techniques of chaos physics and the ranges in which microbubble has stable behavior has been shown by diagrams and also been tabulated to show stability limits of the microbubble, which is extremely important in applications. Results of the radial motion of a polymer-shelled agent display that Qin-Ferrara model which reported in this paper is capable of capturing the essential features of the drug and gene delivery applications. The comprehension of UCA behavior is indispensable to improve its diagnostic and therapeutic implementations in which the nonlinearities cannot be prevented. Nonlinear oscillations of encapsulated microbubble immersed in blood or tissue are scrutinized. The complex dynamics of the microbubble is examined in high acoustic pressure amplitudes with the great magnitude of pulsations which is prevalently utilized in drug and gene delivery applications. The effects of several significant parameters on the behavior of the agent are shown for a wide range of variations which has not been inspected previously. These results provide an exact and comprehensive insight into the system dynamics versus a spacious domain of control parameters. Consequently, we can select a corresponding control process to match our physical conditions. By focusing on the mechanisms governing the transition from the chaotic oscillations to the stable region, this study opens a new horizon in studying  the chaotic behavior of nonlinear dynamics of  a shelled gas bubble submerged in soft tissue or blood.}

  \section*{References}






\end{document}